\documentclass[useAMS,usenatbib]{mn2e}
\usepackage[final]{graphics}
\usepackage{epsfig}
\usepackage{url}
\usepackage{tabu}
\usepackage{threeparttable}

\newcommand{\hel}[2] {He\,{\sc #1}~$\lambda$#2}
\newcommand{\kms}{\mbox{$\mathrm{km~s^{-1}}$}}

\newcommand{\Ion}[2]{#1\,{\sc #2}}

\newcommand{\Ha}{\mbox{${\mathrm H\alpha}$}}
\newcommand{\Hb}{\mbox{${\mathrm H\beta}$}}

\newcommand{\Msun}{\mbox{$\rm M_{\rm \odot}$}}
\newcommand{\Rsun}{\mbox{$\rm R_{\rm \odot}$}}
\newcommand{\tmavg}{\langle \dot M \rangle}
\newcommand{\flux}{\mathrm{erg~cm}^{-2}~\mathrm{s}^{-1}}
\newcommand{\lum}{\mathrm{erg~s}^{-1}}

\newcommand{\fluxd}{\mathrm{erg~cm}^{-2}~\mathrm{s}^{-1}~\mathrm{\AA}^{-1}}
\newcommand{\xmm}{\textit{XMM-Newton}}
\newcommand{\nh}{\mathrm{cm}^{-2}}

\def\hs{\mbox{HS\,0220+0603}}

\def\ergcm{\hbox{erg cm$^{-2}$ s$^{-1}$ }}

\def\lesssim{\mathrel{\hbox{\rlap{\hbox{\lower4pt\hbox{$\sim$}}}\hbox{$<$}}}}
\def\gtrsim{\mathrel{\hbox{\rlap{\hbox{\lower4pt\hbox{$\sim$}}}\hbox{$>$}}}}
\def\Nh{$N_{\rm H}$}
\def\chired{$\chi^{2}_{\nu}$}
\def\chis{$\chi^{2}$}
\def\lx{$L_\mathrm{X}$}
\def\Fx{$F_\mathrm{X}$}

\title[The low state of \hs]{Dynamical masses of a nova-like variable on the edge of the period gap}
\author[P. Rodr\'\i guez-Gil et al.]{P. Rodr\'\i guez-Gil$^{1,2,3}$\thanks{E-mail:prguez@iac.es}, T. Shahbaz$^{1,2}$, T. R. Marsh$^{4}$, B. T. G\"ansicke$^{4}$,
\newauthor
D. Steeghs$^{4}$, K. S. Long$^{5}$, I. G. Mart\'\i nez-Pais$^{1,2}$, M. Armas Padilla$^{1,2}$\thanks{CEI Canarias: Campus Atl\'antico tricontinental},
\newauthor
R. Schwarz$^{6}$, M. R. Schreiber$^{7}$, M. A. P. Torres$^{8}$, D. Koester$^{9}$, V. S. Dhillon$^{10}$,
\newauthor
J. Castellano$^{11}$, D. Rodr\'\i guez$^{12}$\\
$^{1}$Instituto de Astrof\'\i sica de Canarias, V\'\i a L\'actea s/n, La Laguna, E-38205, Tenerife, Spain\\
$^{2}$Departamento de Astrof\'\i sica, Universidad de La Laguna, La Laguna, E-38206, Tenerife, Spain\\
$^{3}$Harvard-Smithsonian Center for Astrophysics, 60 Garden St., Cambridge, MA 02138, USA\\
$^{4}$Department of Physics, University of Warwick, Coventry CV4 7AL, UK\\
$^{5}$Space Telescope Science Institute, 3700 San Martin Drive, Baltimore, MD 21218, USA\\
$^{6}$Leibniz-Institut f\"ur Astrophysik Potsdam (AIP), An der Sternwarte 16, 14482, Potsdam, Germany\\
$^{7}$Departamento de F\'\i sica y Astronom\'\i a, Universidad de Valpara\'\i so, Avenida Gran Breta\~na 1111, Valpara\'\i so, Chile\\ 
$^{8}$SRON, Netherlands Institute for Space Research, Sorbonnelaan 2, NL-3584 CA Utrecht, the Netherlands\\
$^{9}$Institut f\"ur Theoretische Physik und Astrophysik, University of Kiel, 24098 Kiel, Germany\\
$^{10}$Department of Physics and Astronomy, University of Sheffield, Sheffield S3 7RH, UK\\
$^{11}$Observatorio Rodeno (MPC 939), Spain\\
$^{12}$Observatorio Guadarrama (MPC 458), Madrid, Spain}
\begin{document}
\date{Accepted 2015. Received 2015}
\pagerange{} \pubyear{2015}
\maketitle
\begin{abstract}
We present the first dynamical determination of the binary parameters of an eclipsing SW Sextantis star in the 3--4 hour orbital period range during a low state. We obtained time-resolved optical spectroscopy and photometry of \hs\ during its 2004--2005 low brightness state, as revealed in the combined SMARTS, IAC80 and M1 Group long-term optical light curve.

The optical spectra taken during primary eclipse reveal a secondary star spectral type of M5.5$\pm$0.5 as derived from molecular band-head indices. The spectra also provide the first detection of a DAB white dwarf in a cataclysmic variable. By modelling its optical spectrum we estimate a white dwarf temperature of $30\,000 \pm 5\,000$ K.

By combining the results of modelling the white dwarf eclipse from ULTRACAM light curves with those obtained by simultaneously fitting the emission- and absorption-line radial velocity curves and $I$-band ellipsoidal light curves, we measure the stellar masses to be $M_1=0.87 \pm 0.09$\,\Msun\ and $M_2=0.47 \pm 0.05$\,\Msun\ for the white dwarf and the M dwarf, respectively, and an inclination of the orbital plane of $i \approx 79^\mathrm{o}$. A radius of $0.0103 \pm 0.0007$\,\Rsun\ is obtained for the white dwarf. The secondary star in \hs\ is likely too cool and undersized for its mass.

\end{abstract}


\section{Introduction}
\label{sec_intro}

Cataclysmic variables (CVs) are binary systems that contain late-type (K--M) companion stars that fill their Roche lobes and therefore transfer mass to their white dwarf (WD) primary stars. Nova-like (NL) variables are CVs for which a nova or dwarf nova outburst has never been observed. The orbital period distribution of CVs shows a significant paucity of systems in the $\sim 2$--3 hour range, the so-called period gap \citep[see e.g.][and references therein]{gaensickeetal09-1}. CVs with orbital periods above the period gap are expected to have M-dwarf secondary stars up to a period of about $\sim 5.8$ h, while secondary stars of earlier type are to be found for longer orbital periods \citep{knigge06-1}. It is also just above the period gap, between 3 and 4 hours, where a large population of CVs with extreme behaviour is found \citep{thorstensenetal91-1,rodriguez-giletal07-1}. There is mounting evidence that the NLs that populate this orbital period interval harbour intrinsically very bright accretion discs and very hot WDs \citep{townsley+bildsten03-1,araujo-betancoretal05-2,townsley+gaensicke09-1} that current evolutionary theories are unable to account for. In addition, at least 50 per cent of them belong to the SW Sextantis class \citep{rodriguez-giletal07-1}. Their large intrinsic accretion luminosities completely outshine both the WD and its companion most of the time, making direct observations and accurate binary parameter measurements impossible \citep{ciardietal98-1}. However, these CVs are occasionally caught fading towards states of greatly diminished brightness, or `low states'. During the low states, these CVs are 3--5 mag fainter than their normal high state. They can stay at this level for days, months or even years before returning to the high state. Low states appear to be a feature of several types of CV and, in addition, they seem to occur independently of the magnetic field of the WD: they are observed in practically all known strongly magnetic systems without discs (polars), some intermediate polars, and many weakly-magnetic CVs, including a large fraction of NLs and a number of dwarf novae \citep*[e.g.][]{schreiberetal02-1,manser+gaensicke14-1}. 

Although there is broad agreement that low states are the consequence of a reduction in the mass transfer rate from the donor star, the exact mechanism is unclear.\cite{livio+pringle94-1} and \cite{king+cannizzo98-1} proposed an accumulation of large starspots close to the inner Lagrangian point (L1) as a way of inhibiting Roche-lobe overflow, but \cite{howelletal00-1} put the case of the polar ST LMi forward to challenge this scenario and proposed that it is changes in the level of magnetic activity of the secondary star that may be behind the observed high-/low-state transitions. This idea had been already explored by \cite{bianchini92-1}, who invoked solar-like magnetic cycles with time scales of about a decade.

In the low state, only gas provided by the magnetic activity or stellar wind of the donor star would then be available for accretion \citep*[e.g.][]{hessmanetal00-1}. In this respect, entanglement of the magnetic fields of both stars has been proposed to explain the line emission patterns observed in polars during low states \citep[e.g.][]{kafkaetal08-1,masonetal08-1}.



The study of the CV population with orbital periods between 3 and 4 hours is of essential interest to the development of the theory of CV evolution. Particularly, those NL systems closer to the upper boundary of the period gap at an orbital period of about 3.18 hours \citep[a limit set observationally by][]{knigge06-1}. According to the standard stellar structure codes \citep[e.g.][]{chabrier+baraffe97-1}, the M-dwarf secondary stars in these CVs are expected to undergo a major change in their internal structure, developing into a totally convective configuration at a mass of $\sim 0.35~\Msun$. The exact value of this boundary depends on the choice of several parameters, such as the metallicity, opacities and mixing length \citep[e.g.][]{chabrier+baraffe97-1}. We can find in the literature values ranging from 0.27 \citep{gabriel69-1} to 0.4 M$_\odot$ \citep{cox+giuli69-1}. For low-mass stars \cite{reiners+mohanty12-1} place 0.3 M$_\odot$ as the fully-convective limit. In the case of detached white dwarf/main sequence (WDMS) binaries a drop in the fraction of post-common envelope binaries (PCEBs) relative to wide binaries is observed at a mass of $\gtrsim 0.25~\Msun$, which is an independent observation in support of the disrupted magnetic braking mechanism invoked to reproduce the observed orbital period gap in CVs \citep{schreiberetal10-1}. On the other hand, in the population
synthesis study of CVs by \cite{howelletal01-1} their zero-age main sequence models have a fully convective configuration at a mass of $\sim 0.34~\Msun$, while the semi-empirical donor star sequence for CVs of \cite{knigge06-1} suggests a (fully convective) donor mass of 0.2 M$_\odot$ at the upper edge of the period gap. These authors warn against the lack of fundamental input in the form of accurate masses, especially in the 3--4 h orbital period range. In fact, the donor masses used in this period interval come from \textit{indirect} methods such as the mass ratio ($q$)--superhump period excess ($\varepsilon$) relation refined by \cite{pattersonetal05-3}, and the a priori assumption of a constant WD mass of 0.75 M$_\odot$. Moreover, the system used to refine the fit to the $q$--$\varepsilon$ data, thought to have the largest value of the superhump excess ($\varepsilon=0.094$), is BB Doradus, \citep{rodriguez-giletal12-1, schmidtobreicketal12-1} for which an accurate determination of its stellar masses is unavailable.

Another essential observational input to the theory is the time-averaged mass transfer rate, that has to be compared with the predicted secular mass transfer rates \citep*[see][]{kniggeetal11-1}. A medium-term ($10^3-10^5$ yr) mass transfer rate may be derived from measurements of the effective temperature and mass of the accreting WD \citep{townsley+bildsten03-1,townsley+gaensicke09-1}, which again emphasises the need for accurate binary parameters. 

In brief, no direct measurement of the stellar masses in any SW Sex star in the 3--4 h orbital period range exists thus far. However, the CV evolution theory demands appropriate input in the form of dynamical mass solutions and other binary parameters. This is especially true for these CVs close to the upper boundary of the period gap, in which the donor stars may be about to experience the predicted major structural change in their stellar interiors.


In this paper we present the first attempt to measure the fundamental parameters of a SW Sex star in the 3--4 h period range, the eclipsing system HS\,0220+0603 \citep{rodriguez-giletal07-1}, with no initial assumptions about the binary system.


\begin{figure}
\begin{center}
\includegraphics[width=8.45cm]{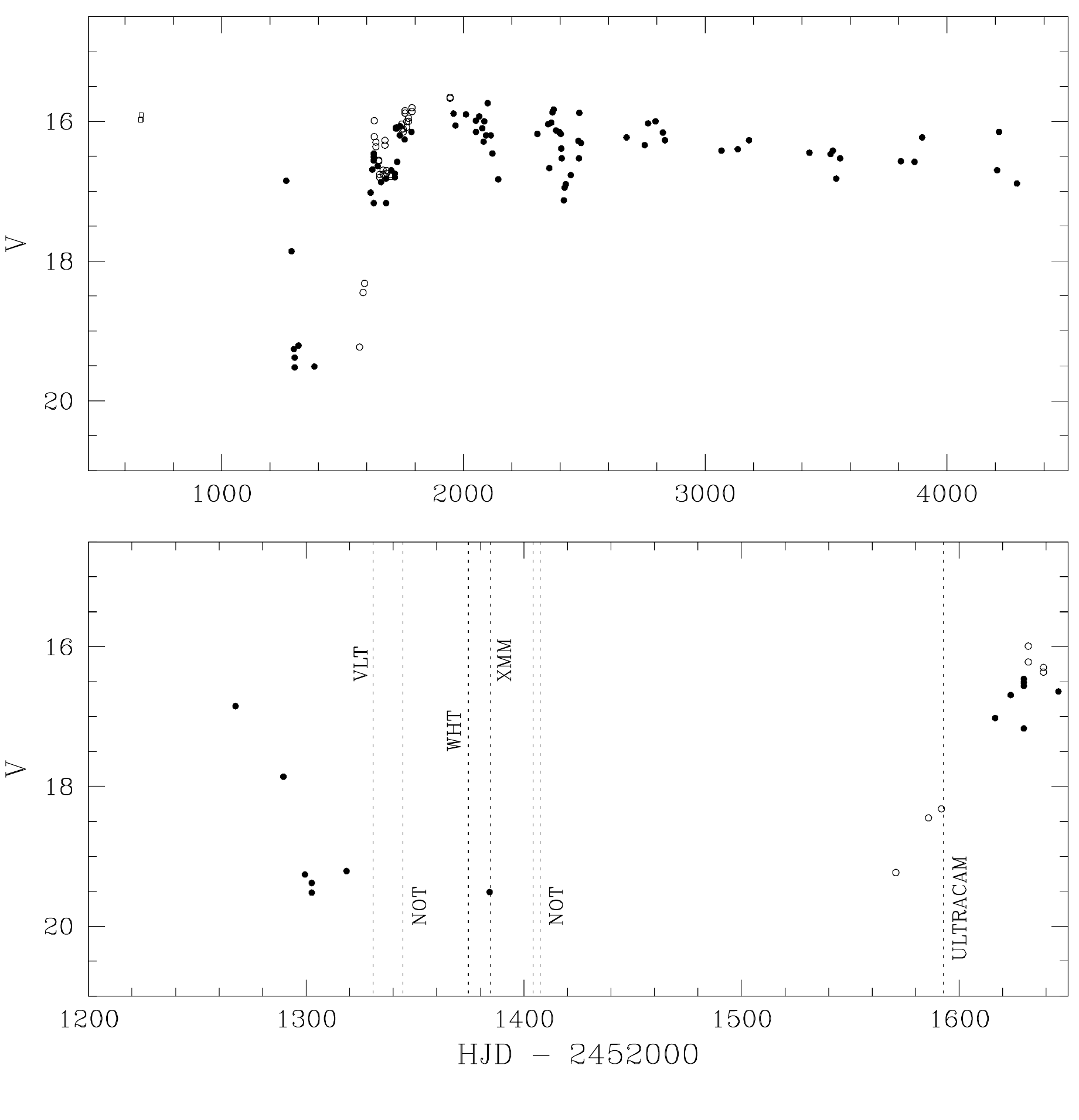}
\end{center}
\caption{Top panel: Long-term, $V$-band light curve of \hs~showing 9.92 years of photometry from 2003 Jan 26 to 2012 Dec 26. Data from the Spanish amateur astronomers of the M1 Group (filled circles), the 1.3-m SMARTS telescope on Cerro Tololo (open circles) and the 0.82-m IAC80 telescope on Tenerife (open squares; first two data points) are shown. Data taken during eclipse are also included in the plot. Bottom panel: Zoomed version of the top panel showing the low state data only. The vertical dashed lines mark the dates of the spectroscopic, photometric and XMM UV observations detailed in Section~\ref{data}.}
\label{fig_longterm_phot}
\end{figure}

\section{Long term light curve}

The long-term $V$-band light curve of \hs~is presented in Fig.~\ref{fig_longterm_phot}. The photometry spans 9.92 years and comprises data provided by Spanish amateur astronomers of the M1 Group, the 1.3-m SMARTS telescope on Cerro Tololo, Chile (operated by the SMARTS Consortium), and the 0.82-m IAC80 telescope at the Observatorio del Teide on Tenerife, Spain. The apparent $V$-band magnitudes were computed relative to the comparison star USNO--A2.0 0900--00554112. Conversion of its $B_\mathrm{A2.0}$ and $R_\mathrm{A2.0}$ magnitudes into the Landolt standard system\footnote{See \url{http://www.pas.rochester.edu/~emamajek/memo_USNO-A2.0.html}.} resulted in a $V$-band magnitude of 12.3. The nearby star USNO--A2.0 0900--00553941 ($V=16.5$) was used as the check star.

The M1 Group reported \hs~entering a low state on 2004 September 19. The system was at $V=16.8$, nearly a magnitude fainter than the previous IAC80 telescope measurements. Almost a month later, on 2004 October 24, the system settled in the low state at $V=19.5$ after fading at a rate of $\sim 0.08$ mag/day. The egress from the low state occurred shortly before 2005 July 19, when the system was observed at $V=19.2$. It then brightened at a relatively slow rate of $\sim -0.05$ mag/day and reached its high state brightness level at $V \approx 16$ around 2005 September 18. This places a lower limit to the low state length of about 1 year. However, just after 2005 September 18 it started to fade again at a rate of $\sim 0.03$ mag/day. \hs~remained at $V \approx 16.7$ for about 64 days before starting its final rise to the high state at the same rate of $\sim -0.03$ mag/day. All the ingress and egress slopes reported here have to be taken with caution due to the scarce sampling of the long-term light curve, but the occurrence of the second fading looks convincing. Further, during this secondary plateau the system showed occasional brightenings with an amplitude of $\sim 0.5$ mag, similar to those seen in BB Dor in the low state \citep{rodriguez-giletal12-1}. Finally, the post-low state monitoring reveals a steady brightness decline at a rate of $\sim 3 \times 10^{-4}$ mag/day.     

\footnotetext[2]{{\sc iraf} is distributed by the National
Optical Astronomy Observatories.}

\section{Data acquisition and reduction}
\label{data}

\subsection{Optical spectroscopy}\label{sec:opt_spec}

\subsubsection{Very Large Telescope}

We obtained four 600-s spectra with the ESO 8.2-m Very Large Telescope (VLT) at Cerro Paranal on 2004 November 20, shortly after the onset of minimum light with the FORS2 spectrograph on the Antu (UT1) telescope.  For the observations the spectrograph was equipped with the 600RI grism, the GG435 order sorter filter and the MIT/LL $2048 \times 4096$\,pixel CCD detector. With a 0.7--arcsec slit, this delivered a spectral resolution of 3.5\,\AA\ (full-width at half-maximum, FWHM) over the $5015-8325$\,\AA\ wavelength range. For flux calibration we used spectra of the flux standard LTT 377. 

\subsubsection{William Herschel Telescope}

We carried out time-resolved spectroscopy of \hs\ on 2005 January 3 with the 4.2-m William Herschel Telescope (WHT) at the Roque de los Muchachos Observatory on La Palma. The double-armed Intermediate dispersion Spectrograph and Imaging System (ISIS) was fitted with the R600B and the R316R gratings, and the $2048 \times 4200$\,pixel EEV12 and the $2148 \times 4700$\,pixel Marconi CCDs in the blue and the red arm, respectively. This setup and a 1.2-arcsec slit width ($\simeq 1.5$-arcsec seeing conditions) granted access to the wavelength intervals $3700-5010$\,\AA\ and $6110-8930$\,\AA\ at 1.8 and 3.3\,\AA~resolution (FWHM) in the blue and red, respectively. The exposure time was 1200 s and we regularly took spectra of arc lamps in order to account for instrument flexure. A flat field image was obtained after every two science exposures of the target in order to correct for fringing in the red arm CCD.
Flux calibration was performed using spectra of the flux standard Feige 34 taken with the same setup. We used the same spectrophotometric standard star to remove the telluric features. See Table~\ref{obstab} for details of the observations. 



\begin{table}
\caption[]{\label{obstab}Log of optical observations.}
\setlength{\tabcolsep}{1.1ex}
\begin{flushleft}
\begin{tabular}{lcccc}
\hline\noalign{\smallskip}
UT Date & Coverage & Filter/Grating & Exp. & \#\,Frames \\
         &   (h)   &                &   (s) & \\  
\hline\noalign{\smallskip}
\multicolumn{5}{l}{\textbf{VLT, FORS2 spectroscopy}} \\
2004 Nov 20    & 0.54 & GRIS\_600RI  & 600 & 4   \\
\hline\noalign{\smallskip}
\multicolumn{5}{l}{\textbf{WHT, ISIS spectroscopy}} \\
2005 Jan 03    & 3.92  & R600B/R316R & 1200/1200 & 12/12 \\
\hline\noalign{\smallskip}
\multicolumn{5}{l}{\textbf{NOT, ALFOSC photometry}} \\
2004 Dec 04     & 3.62 &  $I\#11$  & 40, 60 & 181   \\
2005 Feb 02    & 2.22 &  $I\#11$  & 30, 40, 60 & 107  \\
2005 Feb 05    & 2.35 &  $I\#11$  & 40, 80 & 142   \\
\hline\noalign{\smallskip}
\multicolumn{5}{l}{\textbf{WHT, ULTRACAM photometry}} \\
2005 Aug 09    & 4.27  & $u^\prime$, $g^\prime$, $i^\prime$ & 4.95 & 3089 \\
2005 Aug 10    & 3.93  & $u^\prime$, $g^\prime$, $i^\prime$ & 4.95 & 2841 \\
\noalign{\smallskip}\hline
\end{tabular}
\end{flushleft}
\end{table}

We reduced both the VLT and WHT spectroscopic data using standard procedures in
{\sc iraf}\footnotemark. After subtracting the bias level,
the images were divided by an average flat field that was normalised
by fitting Chebyshev functions of high order to remove the detector
specific spectral response. We then subtracted the sky contribution and optimally-extracted the final spectra following the method
described by \citet{horne86-1}. For wavelength calibration, we fitted
a low-order polynomial to the pixel-wavelength arc data. In calibrating the WHT data, we obtained the wavelength solution for each target spectrum by interpolating between the two nearest arc spectra. All subsequent
analysis was performed with the {\sc molly}\footnote{\url{http://deneb.astro.warwick.ac.uk/phsaap/software/molly/html/INDEX.html}} package. All the spectra were corrected for interstellar reddening using a colour excess of $E(B-V)=0.047$, corresponding to $A_V=0.145$ mag \citep[from the extinction map of][]{schlafly+finkbeiner11-1}.

\subsection{Optical photometry}

\subsubsection{NOT $I$-band light curves}

We used the Andaluc\'\i a Faint Object Spectrograph and Camera (ALFOSC) equipped
with the $2048 \times 2048$ pixel EEV chip (CCD \#8) on
the 2.56-m Nordic Optical Telescope (NOT) on La Palma to obtain time-resolved, 
$I$-band photometry of \hs~on 2004 December 4 and 2005 February 2 and 5. The exposure time ranged between 30 and 80 s due to changing sky transparency conditions during the nights. We used {\sc iraf} to reduce the individual images in a standard way, and to perform aperture photometry on the target and two comparison stars. Differential light curves of the target relative to the main comparison star (C$_1$, USNO--A2.0 0900--00554015) were then computed. We confirmed the stability of C$_1$ by differential photometry relative to the second comparison star. In converting the differential $I$-band magnitudes of \hs~to the standard system, we calculated the $I$-band magnitude of C$_1$ from its Sloan Digital Sky Survey (SDSS) $r$, $i$, and $z$ magnitude measurements (Data Release 7). We then transformed the SDSS magnitudes into $I$-band magnitudes by using the two formulae given by R. Lupton in 2005\footnote{\url{http://www.sdss3.org/dr8/algorithms/sdssUBVRITransform.php}}. The average value for C$_1$ is $I=15.565 \pm 0.007$.

\subsubsection{WHT/ULTRACAM fast photometry}

We also observed \hs~with ULTRACAM on the WHT on La Palma on 2005 August 9 and 10. ULTRACAM is a high-speed, triple-beam CCD camera \citep{dhillonetal07-1} capable of acquiring simultaneous images in three different photometric bands. For \hs~we used the SDSS $u^\prime$, $g^\prime$, and $i^\prime$ filters (see Table~\ref{obstab}). The exposure time was fixed to 4.95 s and the dead time between consecutive exposures was $\sim 24$ ms. We reduced the data with the ULTRACAM pipeline software. Debiassing, flat-fielding and sky subtraction were performed in the standard way. We determined the flux of the source and two close comparison stars with aperture photometry using a variable aperture scaled according to the FWHM of the stellar profile. Note that the ULTRACAM observations took place when the system's brightness was starting to increase (see Fig.~\ref{fig_longterm_phot}), and it was about one magnitude brighter than the low state level. 

\subsubsection{Updated eclipse ephemeris}\label{sec_ephem}

We combined the times of mid-eclipse of the WD in \hs~in the high state reported by \cite{rodriguez-giletal07-1} with new WD mid-eclipse times measured from the light curves in the low state presented in this work to update the eclipse ephemeris. The resulting linear ephemeris is:
$$T_0(\mathrm{HJD})=2\,452\,563.57399(12)+0.149207749(75)\times E~.$$
We used this ephemeris to calculate the orbital phase for all the data presented in this work.

\section{XMM-Newton X-ray and UV observations} \label{sec_xmm}

The \xmm\ satellite \citep{jansenetal01-1} pointed to \hs\ for 30.7~ks on 2015 January 13 (observation ID 0212480101).
The European Photon Imaging Camera (EPIC) consists of two MOS cameras \citep{turneretal01-1} and one PN camera \citep{struderetal01-1}, which were operated in full-frame window imaging mode with the thin optical blocking filter applied. We used the Science Analysis Software (SAS, v.14.0) to carry out the data reduction and obtain the scientific products. 
The observation was affected by episodes of background flaring, therefore a total of 14.2~ks had to be neglected in our analysis. We extracted events using a circular region with a radius of 15 arcsec centred on the source position (RA = 02:23:01.6, Dec. = +06:16:49.6, J2000), and a circular region with a radius of 30 arcsec covering a source-free part of the CCDs to extract background events. The average 0.2--12~keV source net count rate was (8.2 $\pm$ 1.0)$\times 10^{-3}$~count s$^{-1}$ for the PN camera. A very limited number of photons were recorded with the MOS cameras, which did not allow us to perform a spectral analysis. 

We generated the spectrum and the light curve of the PN data, as well as the response files, following the standard analysis threads\footnote{\url{http://xmm.esac.esa.int/sas/current/documentation/threads/}}. The spectral data were grouped to contain a minimum of 10 photons per bin and fit in the 0.3--10~keV energy range using XSpec \citep[v.12.8;][]{arnaud96-1}. We assumed a total Galactic \Ion{H}{i} column density (\Nh) of \Nh=$(3.2\pm 0.1) \times 10^{20}~\nh$ calculated with the \Nh\--$A_{\rm V}$ relation \citep{guver+ozel09-1} using the $A_{\rm V}$ obtained in Section~\ref{sec:opt_spec}. This is consistent with the \Nh=$4.8 \times 10^{20}~\nh$ in the direction to the source found by \cite{kalberlaetal05-1}.  

\begin{figure}
\begin{center}
\includegraphics[width=\columnwidth]{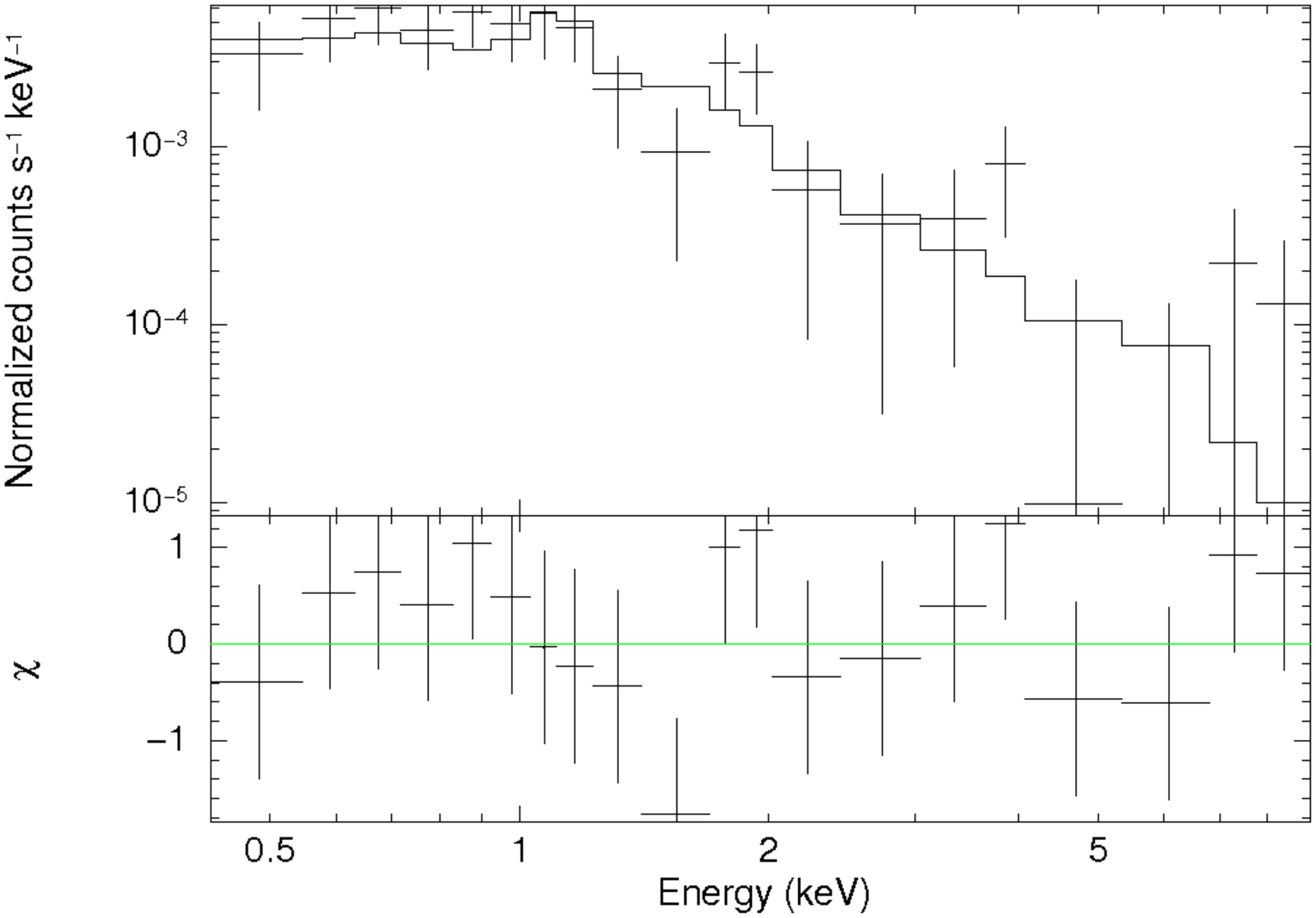}\\
\caption{The 0.3--10 keV PN X-ray spectrum (top) and residuals (bottom). The solid line represents the best fit with a one-temperature thermal plasma model (PHABS*APEC).}
\label{fig:spec}
\end{center}
\end{figure}

Since \hs\ is in a low state, we assume that the bulk of the X-ray emission comes from the M-dwarf companion star. Therefore, we modelled the spectrum with an absorbed one-temperature thermal plasma model, which is typically used for stellar emission. To this end, the plasma thermal emission APEC code \citep{smithetal01-1} was applied. A good fit with \chired=0.7 for 18 degrees of freedom (dof) and p-value \textgreater 0.8 (see Fig.~\ref{fig:spec}) was achieved. The value of the obtained temperature is $kT= 2.9$~keV ($\sim34$~MK)  and the emission measure (EM) is $8.98\times10^{52}~\rm cm^{-3}$. The 0.5--10~keV absorbed (unabsorbed) flux is $2.0\times10^{-14}~\flux$ ($2.1\times10^{-14}~\flux$). Assuming a distance of 780~pc (see Section~\ref{sec_tsh_model}), the corresponding X-ray luminosity is $14\times10^{29}~\lum$ (see Table~\ref{tab:results} for spectral results and X-ray luminosities for different energy ranges in order to compare with the literature).

Our results are comparable with those obtained for single M-dwarf stars \citep{schmitt+liefke04-1}, although in our case the luminosity is higher ($\sim$1 order of magnitude). For example, LMC 335 and 2XMM J043527.2--144301 have temperatures of $kT= 1.22$~keV and $kT= 1.22$~keV, and X-ray luminosities of $1.7\times10^{29}~\lum$ (0.1--7~keV) and $5\times10^{27}~\lum$ (0.2--2~keV), respectively (\citealt{tsangetal12-1}; \citealt*{guptaetal11-1}). When placing \hs\ in the \lx\ vs rotation period plot by \cite{pizzolatoetal03-1} \citep*[also see][]{cooketal14-1}, we see that the luminosity is also higher than rapidly rotating active stars. We therefore cannot discard the presence of X-ray emission due to very low-level accretion.


Simultaneous to the X-ray measurements, observations with the optical/UV monitor telescope \citep[OM;][]{masonetal01-1} were carried out in the imaging mode. The near-ultraviolet \textit{uvw}1 (245--320 nm) and \textit{uvm}2 (205--245 nm) filters were used during $\sim$4.4~ks each. The data were reduced using the {\tt omichain} task. The resulting unreddened fluxes for the \textit{uvw}1 and \textit{uvm}2 bands were $(0.24\pm0.01)\times10^{-15}~\fluxd$, and $(0.48\pm0.02)\times10^{-15}~\fluxd$, respectively (see bottom panel of Fig.~\ref{fig_wd_uv}).

\section{Spectral classification}\label{sec_spec_model}

\subsection{Spectral energy distribution}\label{sec_sed}

\begin{figure}
\begin{center}
\includegraphics[width=8.45cm]{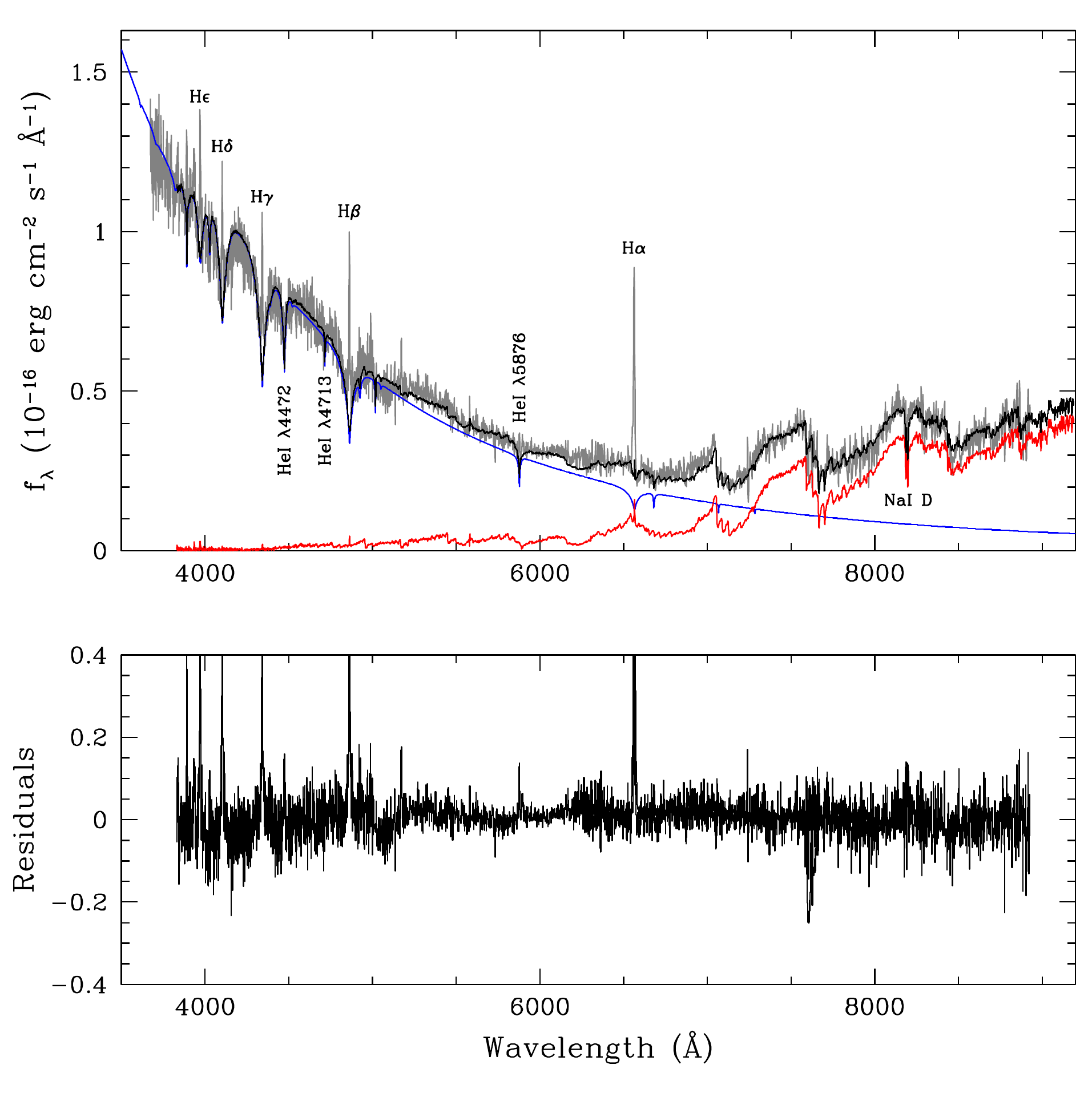}
\end{center}
\caption{Average VLT/FORS and WHT/ISIS spectrum of \hs. Only spectra taken shortly before or after primary eclipse were used in order include the WD spectrum while at the same time minimising the effect of irradiation from the WD on the M-dwarf companion. The data (grey) are fitted with WD (blue)+M-dwarf (red) templates. The solid black line is the best composite fit (see text for details).}
\label{fig_sp}
\end{figure}

The optical spectrum of \hs~in the low state is dominated by the two stellar components. In Fig.~\ref{fig_sp} the contributions of a WD and a mid-M dwarf are apparent in the blue and the red, respectively. The spectral modelling presented below does not require the presence of an accretion disc. In addition, the U-shaped WD eclipse seen in the light curves (see Figs.~\ref{fig_g_eclipse} and \ref{fig_fits}) points to a bare WD or, at most, a WD surrounded by a cold, remnant disc. The emission lines do not reveal a disc either, unlike DW UMa in the low state \citep*{dhillonetal94-1}.   


\begin{table}
\caption{Best fit to the 0.3--10~keV PN spectrum using \textsc{phabs*apec}.}
\label{tab:results}
\begin{threeparttable}
\begin{tabular}{l c c c}
\hline\noalign{\smallskip} 
\multicolumn{4}{c}{\textsc{Parameters model}}\\ 
\Nh & 3.2 $\times10^{20}~\nh$  & & \\ 
$kT$ & $2.9 $~keV & &\\ 
EM$^{a,b}$ & $8.98\times10^{52}~\rm cm^{-3}$ & &\\ 
\chis\ (dof) & 0.7 (18) & &\\
\noalign{\smallskip}
\hline
\noalign{\smallskip}
Energy band & \Fx,$_{\rm abs}$ & \Fx,$_{\rm unabs}$ & \lx$^{b}$ \\
            &  \multicolumn{2}{c}{($10^{-14}~\flux$)} & ($10^{29}~\lum$) \\
0.5--10~keV  &         2.0      &       2.1           & 14      \\
0.1--7~keV  &    2.03     &       2.4           & 16      \\
0.5--10~keV  &   1.25      &       1.62           & 10      \\
\noalign{\smallskip}
\hline

\end{tabular}
\begin{tablenotes}
\item[a]{Emission measure EM= $\int n_{e}n_{H} dV=4\pi D^{2}Norm_{\rm APEC}$.}
\item[b]{To calculate the EM and the luminosities $D=780$~pc was assumed for the distance.}
\end{tablenotes}
\end{threeparttable}
\end{table}

Narrow emission lines of hydrogen are superimposed on the WD absorption troughs. For comparison, the intrinsic FWHM of \Ha\ in the low state is 340 \kms\ at orbital phase 0 (defined as inferior conjunction of the secondary star), whilst the \Ha\ emission line has a FWHM of 1390 \kms\ in the high state \citep{rodriguez-giletal07-1}. The \Ion{Ca}{ii} triplet ($\lambda$8498, $\lambda$8542, and $\lambda$8662) in emission is apparent around orbital phase 0.5. Some \hel{i}{5876} emission is also present. We will later show that these narrow emission lines originate on the donor side of the binary system, and are due to irradiation from a hot WD and/or chromospheric emission. The \hel{ii}{4686} and Bowen blend emissions observed in the high state are absent.  

In determining the spectral type of the two stars we used an average spectrum of \hs~computed from spectra corresponding to a limited range of orbital phases around the primary eclipse. This minimises the contribution from the side of the secondary star that faces the WD and yet includes the spectrum of the WD. In particular, we only used VLT/FORS2 spectra obtained at orbital phases 0.15 and 0.20, WHT/ISIS red spectra at 0.92, 0.93, and 0.11, and at 0.82, 0.92, 0.93, 0.11, and 0.21 for the WHT/ISIS blue spectra. Note that we use this average spectrum in the subsequent analyses unless otherwise stated. Also note that this approach will eliminate any information on the spectral type variation with orbital phase, but it is necessary as the signal-to-noise ratio of the individual spectra is far from being optimal for this purpose.

The spectral contribution of the WD is very blue in colour, and contains both Balmer and \Ion{He}{i} lines, making this the first definite detection of a DAB WD in a CV. The composite spectrum superimposed on the average spectrum of \hs\ presented in Fig.~\ref{fig_sp} contains our DAB WD model with temperature and surface gravity of 30\,000 \,K and $\log g= 8.35$, respectively, and a dwarf M5.5 secondary star (see Section~\ref{sec_molecindices}).


\begin{figure}
\begin{center}
\includegraphics[width=8.55cm]{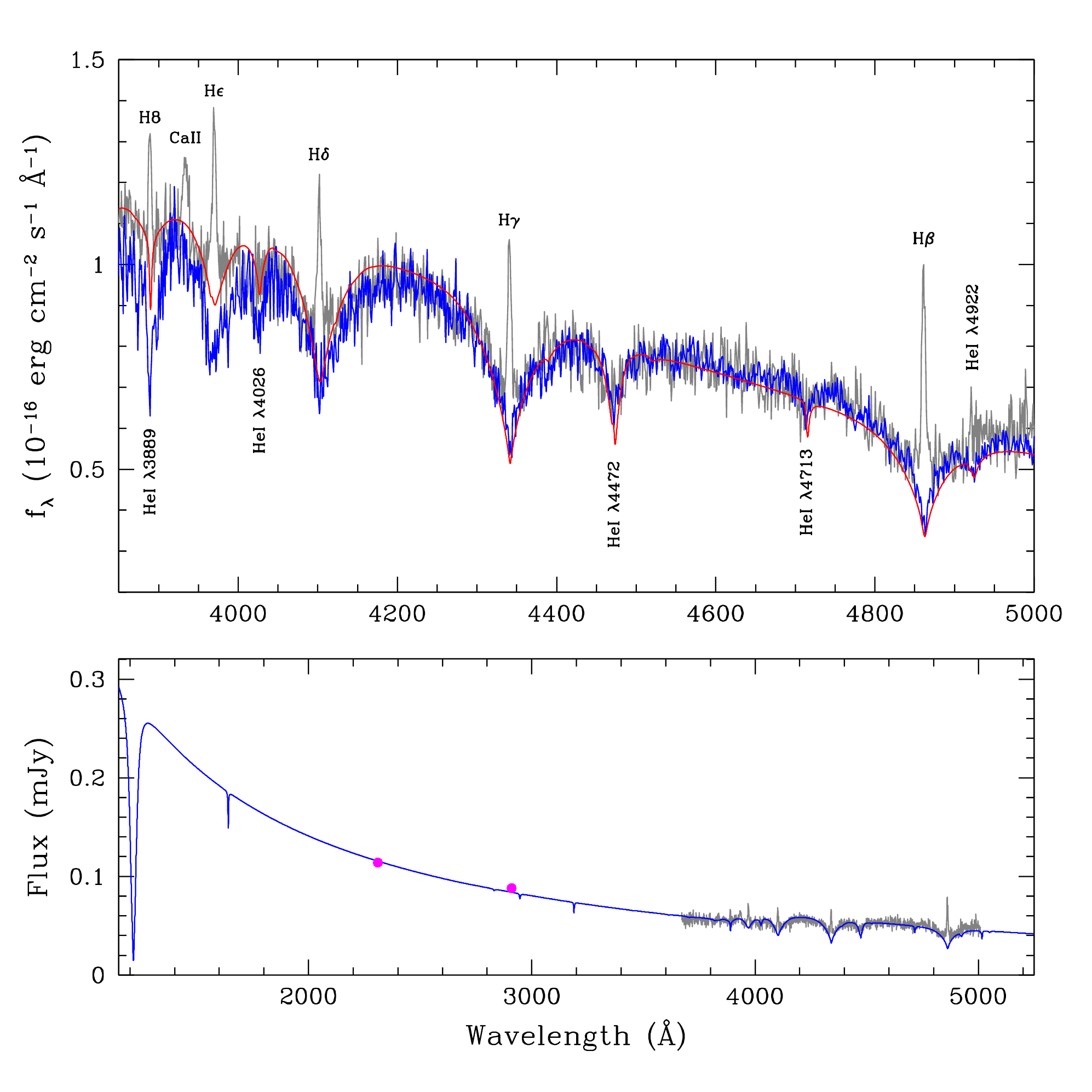}
\end{center}
\caption{Top panel: Average WD spectrum around eclipse (grey) with our DAB WD model (red; $T_\mathrm{eff}=30\,000$ K, $\log g=8.35$) and spectrum of the field DAB WD SDSS\,J084742.22+000647.6 (blue; $T_\mathrm{eff}=30545$ K, $\log g=8.49$) superimposed. Note that the contribution of the secondary star has not been subtracted, so there is a flux excess increasing with wavelength. Bottom panel: Average WD spectrum around eclipse (grey) with our DAB WD model (blue) and the XMM UV flux measurements after correcting for interstellar reddening (circles).}
\label{fig_wd_uv}
\end{figure}

The de-reddened UV fluxes (see Section~\ref{sec_xmm}) are plotted together with our WD model and the average WHT/ISIS blue spectrum in the bottom panel of Fig.~\ref{fig_wd_uv}. The UV observations fit well into the WD model with $T_\mathrm{eff} = 30\,000$ K. For comparison, we show in Fig.~\ref{fig_wd_uv} the SDSS spectrum of a (single) DAB WD with similar atmospheric parameters \citep[SDSS\,J084742.22+000647.6, $T_\mathrm{eff} = 30545 \pm 241$ K, $\log g = 8.49 \pm 0.05$;][]{kleinmanetal04-1} along with the average spectrum of the WD in \hs\ and the DAB model. The SDSS WD and the \hs~spectra show a good match. Given the uncertainties in the flux calibration of the average WD spectrum and the UV fluxes, and considering that the UV and the optical observations were not simultaneous, it is difficult to establish a formal error on the WD temperature, but we conservatively estimate that the quoted value is accurate to within 5000\,K.

\subsection{Secondary star molecular-band indices}\label{sec_molecindices}

In this section we present the determination of the spectral type of the secondary star. \cite{boeshaar76-1} introduced a spectral classification scheme for M dwarfs based on molecular-band ratios, which has also been used for CVs \citep{wade+horne88-1}. The work of \citet*{reidetal95-1} used a similar set of spectral indices to measure the depth of selected molecular band heads. These authors improved on the M-dwarf spectral sequence presented by \citet*{kirkpatricketal91-1} for classification in the red ($6300-9000$\,\AA). These indices have been extensively calibrated against spectral type by many authors \citep*[see][and references therein]{lepineetal03-1}.

An assessment of the spectral type of the secondary star in \hs~can therefore be done by measuring a set of these spectral indices from the average WHT/ISIS red spectrum and comparing them with the indices observed for calibrated M dwarfs. We followed the strategy described by \cite{lepineetal03-1} for mid to late M-dwarfs, focusing on the VO1, VO2, and TiO7 indices (see their table 2 for details). We did not use the TiO6 index as the continuum is defined very close to the $\mathrm{O}_2$ telluric absorption complex at $\sim 7650$\,\AA, and residuals from the telluric correction remain in that region. Prior to this analysis we first removed the contribution of the WD. We present the flux ratios for \hs~and SDSS M-dwarfs ranging from M2 to M8 in Table~\ref{tab_indices}. \cite{lepineetal03-1} present linear functions to obtain the spectral type of M dwarfs from the molecular-band indices. Using their equations (4), (6), and (7) we obtain a spectral type for the secondary star in \hs~of M5.5, M5.6, and M5.4 for the VO1, VO2, and TiO7 indices, respectively. The indices obtained for the SDSS M-dwarf templates (Table~\ref{tab_indices}) also point to a M5--6 spectral type. The linear calibrations of \cite{lepineetal03-1} can provide spectral types accurate to half a spectral type. We can therefore conclude that the M-dwarf secondary star in \hs~shows a low-state spectral type of M5.5$\pm$0.5 when observed close to its inferior conjunction.

\label{sec-results}
\section{Binary parameters}
\subsection{White dwarf eclipse modelling \label{sec_ultracam_fits}}

\begin{figure}
\begin{center}
\includegraphics[width=8.45cm]{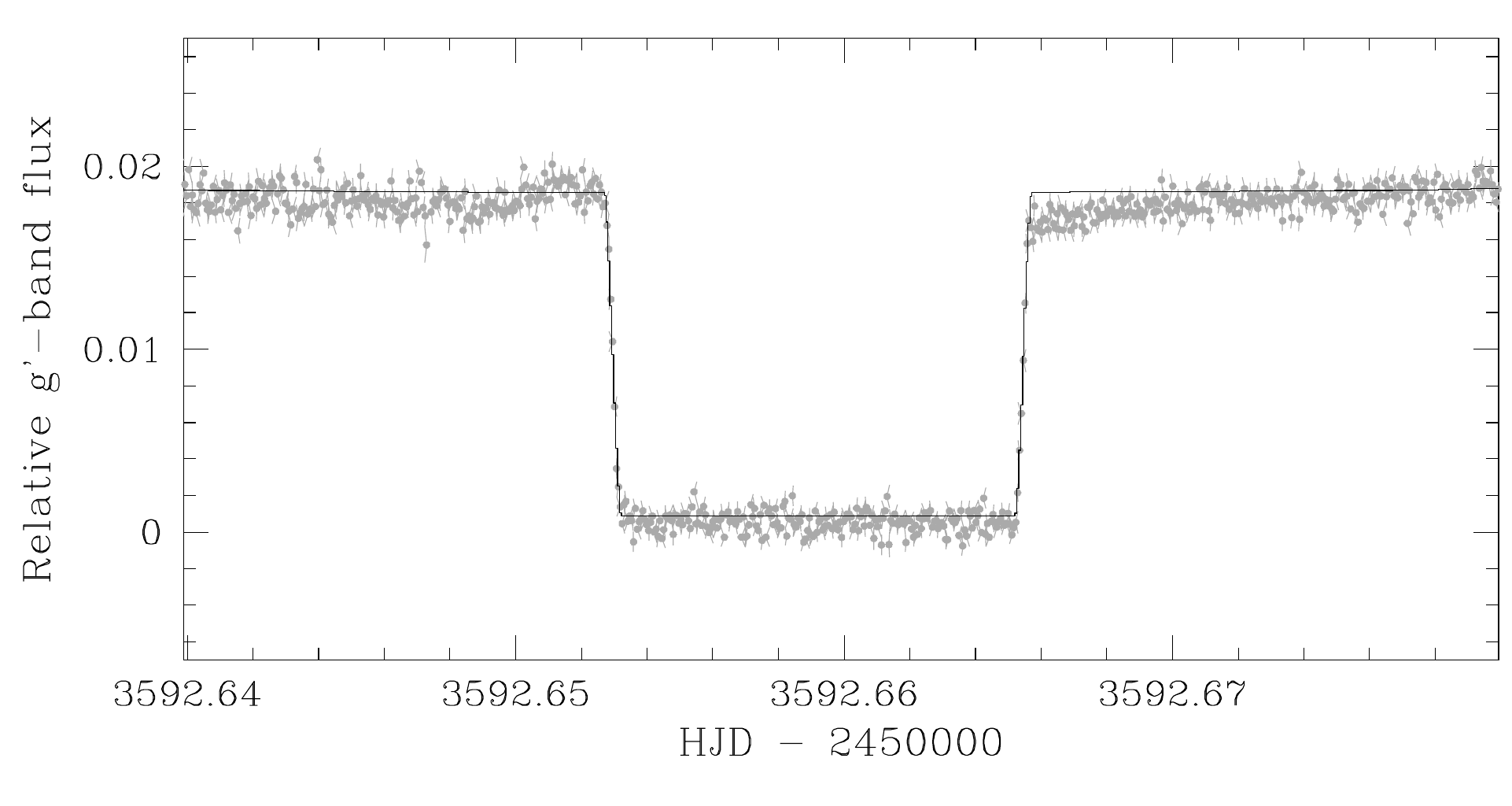}
\end{center}
\caption{Modelling of the WHT/ULTRACAM WD eclipse in the $g^\prime$-band. The data were fitted with the \textsc{lcurve} code. The solid line represents the best fit.}
\label{fig_g_eclipse}
\end{figure}

In order to obtain constraints on the binary parameters we fitted light curve models to the ULTRACAM $g^\prime$-band eclipse data. The $g^\prime$-band light curve has the best quality and the deepest eclipse of the three recorded by ULTRACAM (see Fig.~\ref{fig_g_eclipse}). Of special interest is the mass ratio $q = M_{\rm 2}/M_{\rm 1}$ (where $M_{\rm 1}$ and $M_{\rm 2}$ are the masses of the WD and secondary star, respectively) as a function of the orbital inclination ($i$), that can be tested against an analogous function derived from the joint modelling of the emission- and absorption-line radial velocity curves and the ellipsoidal modulation plus eclipse curves (see Section~\ref{sec_tsh_model}). 


\begin{table}
\caption[]{\label{tab_indices}Molecular band-head indices.}
\setlength{\tabcolsep}{1.1ex}
\begin{center}
\begin{tabu} to \columnwidth {X[c]X[c]X[c]X[c]}
\hline\noalign{\smallskip}
Star & VO1  &  VO2 & TiO7          \\
\hline\noalign{\smallskip} 
\hs&0.875&   0.649& 0.756\\  
M2~V            &0.947&   0.899& 0.956 \\ 
M3~V            &0.920&   0.821& 0.914  \\ 
M4~V            &0.904&   0.758& 0.876  \\
M5~V            &0.874&   0.654& 0.791  \\
M6~V           &0.851&   0.601& 0.762  \\
M7~V           &0.829&   0.537& 0.703  \\
M8~V           &0.799&   0.443& 0.626  \\
\hline\noalign{\smallskip}
\end{tabu}
\end{center}
\end{table}

For the modelling of the ULTRACAM light curve we made use of the {\sc lcurve} code developed by one of us (TRM; see \citealt{copperwheatetal10-1} for a description, and e.g. \citealt{pyrzasetal09-1} for its application to light curves of WD/M-dwarf binaries). The {\sc lcurve} code computes a model from input parameters provided by the user. Model light curves are then fitted to the data using Levenberg--Marquardt minimisation. Both fixed and free parameters can be set by the user, which gives full flexibility to the code operation.

The results of the spectral modelling presented in Section~\ref{sec_sed} allowed us to fix the temperature of the WD to $T_1=30\,000$\,K. We also fixed the limb darkening coefficients to 4--parameter values as listed in \cite{gianninasetal13-1} for $T_1=30\,000$\,K and $\log g=8.5$. The mass ratio $q$, the scaled WD radius $r_1=R_1 /a$ (with $a$ the binary separation) and the effective temperature of the secondary star $T_2$ were left as free parameters. For the scaled secondary star radius $r_2=R_2 /a$, the code calculates the proper non-spherical shape of the secondary star's Roche lobe for each value of $q$. We performed a grid search in $q$, with $q$ ranging from 0.25 to 0.60 in steps of 0.05. The results are presented in Table~\ref{table_tom_fitting}.


\subsection{Simultaneous radial velocity and ellipsoidal modulation modelling}\label{sec_tsh_model}

The NOT $I$-band light curves were phase folded according to the eclipse ephemeris given in Section~\ref{sec_ephem} and averaged into 43 orbital phase bins. To interpret the $I$-band light curve and the absorption- and emission-line radial velocity curves we used the 
{\sc xrbcurve} model described in \cite{shahbazetal03-1}, which has successfully been 
used to model the light curves and radial velocity curves of neutron 
star and black hole X-ray binaries \citep{shahbazetal04-1}. Briefly, the 
model includes a Roche-lobe filling secondary star, the effects of UV/X-ray
heating of the secondary star by a source of high energy photons 
from a compact object, an accretion disc and mutual eclipses of the disc 
and star. In the case of \hs\ in the low state, where the eclipse of the WD by 
the secondary star is clearly seen, we simulate the presence of a bare WD by assuming that it can be described by a blackbody
cylindrical accretion disc (with equal radius and height).

\begin{table}
\caption[]{\label{table_tom_fitting}ULTRACAM WD eclipse modelling parameters.}
\setlength{\tabcolsep}{1.1ex}
\begin{center}
\begin{tabu} to \columnwidth {X[c]X[c]X[c]}
\hline\noalign{\smallskip}
Mass ratio & Inclination & Scaled WD radius\\
$q~(=M_2/M_1)$ & $i$ (deg) &  $r_1~(=R_1/a)$\\
\hline\noalign{\smallskip} 
0.260  &  88.71(15) &   0.0101(5)\\   
0.300  &  85.48(04) &   0.0095(5)\\ 
0.350  &  83.51(03) &   0.0091(4)\\
0.400  &  82.11(03) &   0.0087(5)\\ 
0.450  &  80.99(02) &   0.0083(4)\\ 
0.500  &  80.05(02) &   0.0080(4)\\ 
0.550  &  79.24(02) &   0.0078(4)\\ 
0.600  &  78.52(02) &   0.0075(4)\\
\hline\noalign{\smallskip}
\end{tabu}
\end{center}
\end{table}

The geometry of the binary system is given by the orbital inclination 
$i$, the binary mass ratio $q$, and the Roche-lobe filling factor $f_\mathrm{R_2}$ of the 
secondary star. The light arising from the secondary star depends on its 
mean effective temperature ($T_{\rm 2}$), the gravity darkening exponent 
$\beta$ and the X-ray/UV albedo $W$. The additional light due to UV/X-ray 
heating is given by the unabsorbed UV/X-ray flux ($F_{\rm X,0}$). The light 
from the WD is given by its scaled radius and effective temperature. The distance to the source in parsecs ($d_{\rm pc}$), the 
orbital period $P$ and the radial velocity amplitude of 
the secondary star ($K_{\rm 2}$) set the scale of the system.

We use \textsc{phoenix} model-atmosphere fluxes \citep*{hauschildtetal99-1} to 
determine the intensity distribution on the secondary star and blackbody fluxes for the light arising from the WD. For both cases 
we use a quadratic limb-darkening law with coefficients taken from 
\cite{claret98-1}, to correct the intensity. We assume that the 
secondary star is in synchronous rotation and completely fills its Roche 
lobe. Since the late-type secondary star should have a convective envelope, we 
fix the gravity darkening exponent to 0.08 \citep{lucy67-1}. Its albedo ($W$) is fixed at 0.40 \citep{dejongetal96-1}.

The irradiating flux can strongly alter the distribution of temperature 
across the face of the secondary star and thus the observed optical light curves 
and radial velocities. For the light curves, the heating of the 
secondary star is computed in the same way as described in 
\cite{shahbazetal03-1}, who calculate the increase in unperturbed 
local effective temperature due to the irradiating external source. 
Substantial heating of the secondary star also shifts the effective 
centre of the secondary, weighted by the strength of the absorption and 
emission lines, from the centre of mass of the star. This results in a 
significant distortion of the radial velocity curve leading to
spuriously high or low $K$-amplitudes using absorption or emission 
lines, respectively, arising from the secondary star 
(Phillips, Shahbaz \& Podsiadlowski \citeyear{phillipsetal99-1}; Shahbaz et al. \citeyear{shahbazetal00-1}). 
To model the radial velocity curves we determine the fraction of the surface of the star that contributes 
to the absorption-line or emission-line radial velocity.  
\begin{table}
\centering
\caption{
\label{modvar} List of the key model variables and what they represent.}
\begin{tabular}{@{}l|l@{}}
\hline
Variable   & Definition                                       \\  
\hline
$M_{\rm 1}$   & Mass of the white dwarf                       \\
$M_{\rm 2}$    & Mass of the secondary star                       \\
$q$            & Binary mass ratio defined as $M_2/M_1$       \\
$i$            & Orbital inclination                               \\
$P$  & Orbital period                                   \\
$a$            & Binary separation                                  \\ 
$F_{\rm X,0}$  & Unabsorbed heating flux                            \\
$d_{\rm pc}$  & Distance in pc                                         \\
$K_{\rm 2}$    & Secondary star's radial velocity amplitude  \\
$T_{\rm 2}$    & Secondary star's mean effective temperature      \\
$R_{\rm 1}$   & White dwarf radius                          \\
$T_{\rm 1}$   & White dwarf effective temperature                          \\
$\beta$        & Gravity darkening exponent                       \\
$W$            & Albedo of the secondary star                     \\
$E(B-V)$       & Colour excess                                    \\ 
\hline
\end{tabular}
\end{table}
These are 
described by factors $F_{\rm AV}$ and $F_{\rm EV}$, respectively, and 
represent the fraction of the external radiation flux that exceeds the 
unperturbed flux, e.g. $F_{\rm AV}$=1.0 assumes all the area of the star visible 
by the heating source is irradiated and does not contribute to the 
absorption-line radial velocity, whereas for $F_{\rm EV}$=1.10 only 
surface elements on the star where the external radiation flux is greater than 
10 per cent of the unperturbed flux contribute to the emission-line radial 
velocity (Billington, Marsh \& Dhillon \citeyear{billingtonetal96-1}; Phillips et al. \citeyear{phillipsetal99-1}). 
The line flux of each element on the star is calculated using the equivalent width and continuum flux value, and the equivalent width of elements that do not contribute are set to zero. For ease of reference, the key variables are listed in Table\,\ref{modvar}.

In determining the binary parameters we simultaneously fit 
the photometric $I$-band light curve (43 data points) and the absorption-line (\Ion{Na}{i} doublet) and 
emission-line (\Ion{Ca}{ii} triplet) radial velocity curves (11 and 7 data points, respectively) 
with our model to represent the photometric and radial velocity 
variations.  We used the differential evolution algorithm described in 
\cite{shahbazetal03-1} to fit the data, which is robust and simple.  

The model parameters that determine the shape and amplitude of the 
optical light curves, absorption-line and emission-line radial velocity 
curves are $i$, $q$, $T_{\rm 2}$, $F_{\rm X,0}$, $d_{\rm pc}$, $W$, $R_{\rm 
1}$, $T_{\rm 1}$, $K_{\rm 2}$, $F_{\rm AV}$, and $F_{\rm EV}$.  For the 
fitting procedure there are three extra parameters: the phase 
shift for the optical light curve $\delta \phi_{\rm LC}$, and the phase 
and radial-velocity shift for the absorption-line radial velocity curve 
$\delta \phi_{\rm RV}$ and $\gamma_{\rm RV}$, respectively (the 
absorption-line radial velocity curve allows us to define the phase 0.0, i.e. 
inferior conjunction of the secondary star, and the systemic velocity, 
hence we use the same values for the fit to the emission-line radial 
velocity curve).

We fixed $T_{\rm 2} =3000$\,K for a mid-M dwarf (see e.g. \citealt{leggettetal96-1}), 
a WD blackbody temperature of $T_{\rm 1}=30\,000$\,K, $E(B-V)$ = 0.047 
with the constraint $M_1 < 1.4 $~\Msun\
and then performed a grid search in $i$. Given 
that there are three different data sets with different numbers of 
data points, to optimise the fitting procedure we assigned relative 
weights to them. After our initial search of the 
parameter space, which resulted in a good solution, we scaled the 
uncertainties on each data set (i.e. the light curve and the radial 
velocity curves) so that the total reduced $\chi^2$ of the fit was $\sim 1$ for each data set separately.  
The fitting procedure was then repeated to produce the final set of parameters. 
We have checked that the results do not significantly change upon varying $T_2$ 
by a few hundred degrees.

\begin{figure}
\begin{center}
\includegraphics[width=\columnwidth]{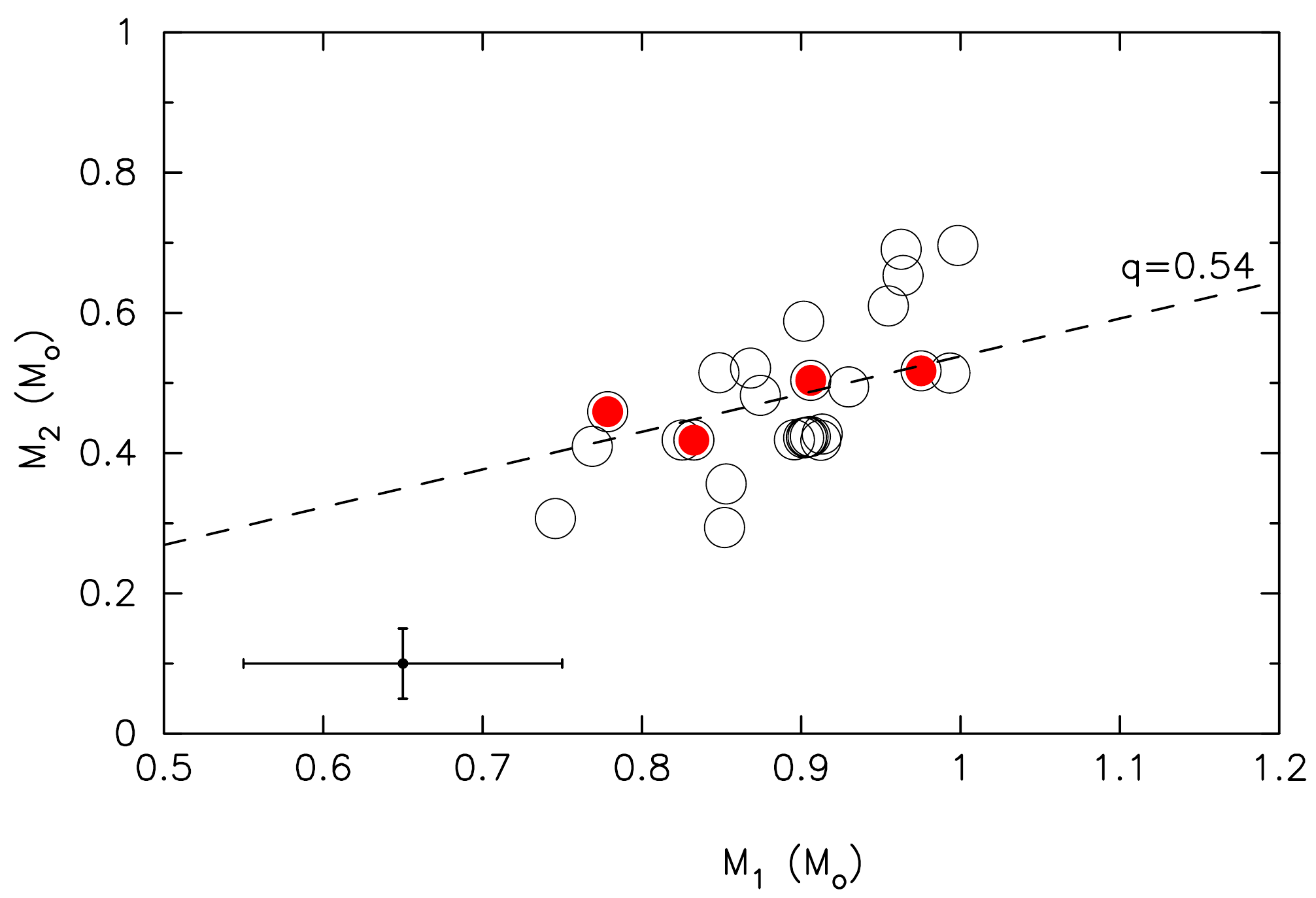}
\end{center}
\caption{$M_1$--$M_2$ plane with the best fit solutions. 
The solid red circles show the 99 per cent confidence level solutions. The dashed line corresponds to a mass ratio of $q=0.54$. The bars at the lower left represent the typical error.}
\label{m1m2_fits}
\end{figure}


By combining the best fit model parameters $q$, $i$ and $K_2$ with the mass function 
equation, we can determine $M_1$ and $M_2$.  In Fig.~\ref{m1m2_fits} we show the best fit 
solutions in the ($M_1 - M_2$) plane, where the most probable value for the binary masses are 
$M_1 = 0.87 $~\Msun\ and $M_2 = 0.47 $~\Msun. The red circles in the figure show the 99 per cent confidence 
level solutions. Table~\ref{modresult} shows the mean and \textit{rms} uncertainties in the parameters 
values, where we have only used solutions with the 99 per cent confidence level. As one can 
see the ($M_1-M_2$) solution allows us to constrain $q$; the dashed line in Fig.~\ref{m1m2_fits} shows the best fit with 
$q=0.54 \pm 0.02$, which when combined with the $(i-q)$ solutions obtained 
from the ULTRACAM eclipse fitting (see Section~\ref{sec_ultracam_fits}), gives $i\sim 79$ degrees.
In Fig.~\ref{fig_fits} we show the light curve and radial velocity curves for the best fit
model parameters; $M_1 = 0.87 $~\Msun\ and $M_2 = 0.47 $~\Msun, $i =79^{\circ}$, $T_{\rm 2} =3000$\,K, and 
$T_{\rm 1}=30\,000$\,K.
 
 
We also present a model prediction of the \Ha~emission-line radial 
velocity curve (not included in the fitting process) assuming that all 
of the inner face of the secondary star is irradiated. The model nicely 
reproduces the observed \Ha~radial velocity curve, but note that the 
\Ha~emission line is also seen around zero phase, indicating that it is 
not entirely driven by irradiation. We suggest that chromospheric 
\Ha~emission is also occurring in the secondary star \citep[also see][for the case of VV Pup in the low state]{masonetal08-1}.


\begin{table}
\caption{\label{modresult} 
Results obtained from the model fits to the $I$-band optical light curve,  \Ion{Na}{i}
absorption-line and \Ion{Ca}{ii} emission-line radial velocity curves 
of \hs\ for $T_{\rm 2}=3000$\,K, and $T_{\rm 1}=30\,000$\,K.}
\setlength{\tabcolsep}{1.1ex}
\begin{center}
\def\arraystretch{1.5}
\begin{tabu} to \columnwidth {X[l]X[c]}
\hline\noalign{\smallskip}
Parameter                 & Value   \\			 
\hline\noalign{\smallskip} 
$q$                        &  $0.54 \pm 0.03$ \\			 
\noalign{\smallskip} 
$K_{\rm 2}$\,(\kms)        &  $284  \pm 11$\\
\noalign{\smallskip} 
$d_{\rm pc}$               &  $740 \pm 26 $\\
\noalign{\smallskip} 
$\log F_{X,0}$\,(\ergcm)   & $-11.7 \pm 0.05$ \\
\noalign{\smallskip} 
$\gamma_{\rm AV}$\,(\kms ) &  $1.8 \pm 2.3$ \\
\noalign{\smallskip} 
$F_{\rm AV}$               &  $1.06 \pm 0.06$ \\
\noalign{\smallskip} 
$F_{\rm EV}$               &  $1.03 \pm 0.02$ \\
\noalign{\smallskip} 
$M_{\rm 1}$\,(\Msun)       &  $0.87 \pm 0.09$  \\
\noalign{\smallskip} 
$M_{\rm 2}$\,(\Msun)       &  $0.47 \pm 0.05$ \\
\hline\noalign{\smallskip}
\end{tabu}
\end{center}
\end{table}


\section{Discussion}
\subsection{The white dwarf} \label{sec_discuss_wd}

The spectral decomposition of the optical spectrum of \hs~in the low state revealed a DAB WD with $T_\mathrm{eff} \simeq 30\,000$ K and $\log g \simeq 8.35$. On the other hand, our dynamical solution of the binary system provided a WD with mass $M_1=0.87 \pm 0.09$\,\Msun. Cubic interpolation of the $q-r_1$ data from the ULTRACAM eclipse modelling (see Table~\ref{table_tom_fitting}) yields a scaled WD radius $R_1/a=0.0078 \pm 0.0004$ for a mass ratio $q=0.54$, which translates into a WD radius $R_1 = 0.0103 \pm 0.0007$ \Rsun. The WD surface gravity derived from the measured mass and radius is $\log g = 8.36$, in good agreement with the $\log g \simeq 8.35$ value found from spectral modelling (Section~\ref{sec_sed}). When the WD radius is used in the WD spectral modelling a distance of $\sim 763$ pc is derived, which is consistent with the results presented in Section~\ref{sec_tsh_model} (see Table~\ref{modresult}). 

\begin{figure}
\begin{center}
\includegraphics[width=4cm]{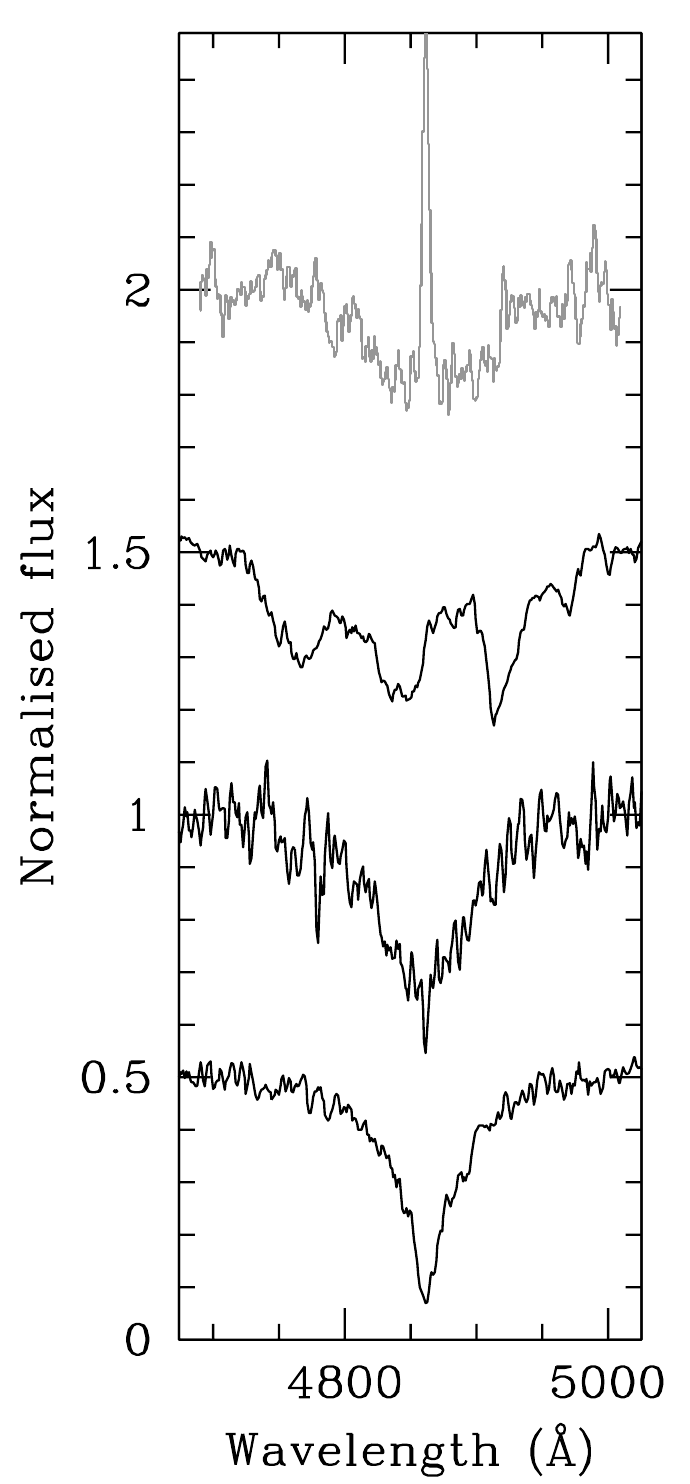}
\end{center}
\caption{Comparison of the \hs\ average \Hb\ absorption profile (top grey spectrum) with the \Hb\ absorption profiles of three magnetic WDs (showing the characteristic Zeeman splitting) with magnetic fields of 8.4, 4.1 and 1.0 MG (from top to bottom; see text for details).}
\label{fig_magnet_comp}
\end{figure}

Circular polarimetry observations of SW Sex stars show that some systems may contain magnetic WDs \citep{rodriguez-giletal01-1,rodriguez-giletal02-1,rodriguez-giletal09-1}. In order to check for any Zeeman splitting in the WD spectrum of \hs\ we compare the average \Hb\ absorption profile of \hs\ with three magnetic WDs (Fig.~\ref{fig_magnet_comp}): SDSS\,J151130.17+422023.00 \citep[$T_\mathrm{eff}=30882$\,K, $B_1=8.4$\,MG,][]{kepleretal13-1}, SDSS\, J154305.67+343223.6 \citep[$T_\mathrm{eff}=25\,000$\,K, $B_1=4.1$\,MG,][]{kulebietal09-1}, and SDSS\,J091005.44+081512.2 \citep[$T_\mathrm{eff}=25\,000$\,K, $B_1=1.0$\,MG,][]{kulebietal09-1}. We can therefore rule out a WD magnetic field in excess of 8 MG in \hs, but $B_1 \lesssim 4.1$\,MG would pass undetected due to the poor signal-to-noise ratio. Blue spectra of better quality are needed to confirm any Zeeman splitting in the low state.



Observations of accreting WDs in CVs indicate that they are significantly more massive than isolated WDs. The statistical studies of the CV population by \cite{smith+dhillon98-1} and \cite{knigge06-1} produced mean WD masses of $0.80 \pm 0.22$ \Msun~(above the period gap, $0.69 \pm 0.13$ \Msun~below) and $0.75 \pm 0.05$ \Msun, respectively. Also, \citet{savouryetal11-1} and \citet*{zorotovicetal11-1} report more massive WDs in CVs. The WD mass for \hs\ fits well within these results. In addition, the WD mass of the eclipsing dwarf nova IP Peg ($P=3.80$\,h) is measured to be $1.16 \pm 0.02$ \Msun\ \citep{copperwheatetal10-1}. On the other hand, the mass distribution of isolated WDs peaks at $\lesssim 0.6$ \Msun\,(e.g. Koester, Schulz \& Weidemann \citeyear{koesteretal79-1}; Bergeron, Saffer \& Liebert \citeyear{bergeronetal92-1}; Kepler et al. \citeyear{kepleretal07-1}; Falcon et al. \citeyear{falconetal10-1}; Tremblay et al. \citeyear{tremblayetal13-1}). 

The distribution of WD effective temperature as a function of the orbital period may serve as a probe of CV evolution. Changes in the mass transfer rate reflect in the WD effective temperature, which is determined by compressional heating of the WD by the accreted matter. Therefore, $T_\mathrm{eff}$ can be a good tracer of the medium-term ($10^3-10^5$ yr) mass transfer rate \citep{townsley+bildsten03-1,townsley+gaensicke09-1}. The WD effective temperatures predicted by the evolutionary sequences of \cite{kniggeetal11-1} are significantly cooler than the data below the period gap, and show a large disagreement in the 3--4 h orbital period range, where the scarce measurements point to much hotter WDs than expected (see fig. 5 of \citealt{townsley+gaensicke09-1} and fig. 16 of \citealt{kniggeetal11-1}). However, the relationship between the time-averaged mass transfer rate and the WD temperature has to be used with caution. As \cite{townsley+gaensicke09-1} point out, the mass transfer rate for a given $T_\mathrm{eff}$ strongly depends on the WD mass, which again highlights the need for independent mass measurements. In addition, a dominant WD in the blue/UV is also a requirement for a reliable $T_\mathrm{eff}$ determination. Hence, studies of the WDs in CVs in the low state like the one presented here are important to test the current CV evolutionary sequences in the 3--4 h regime. In this regard, \cite{kniggeetal11-1} suggest that the large scatter of the WD temperatures observed in this orbital period interval may be the result of the temperatures not tracing the secular mass transfer rates. 

\begin{figure}
\begin{center}
\includegraphics[width=\columnwidth]{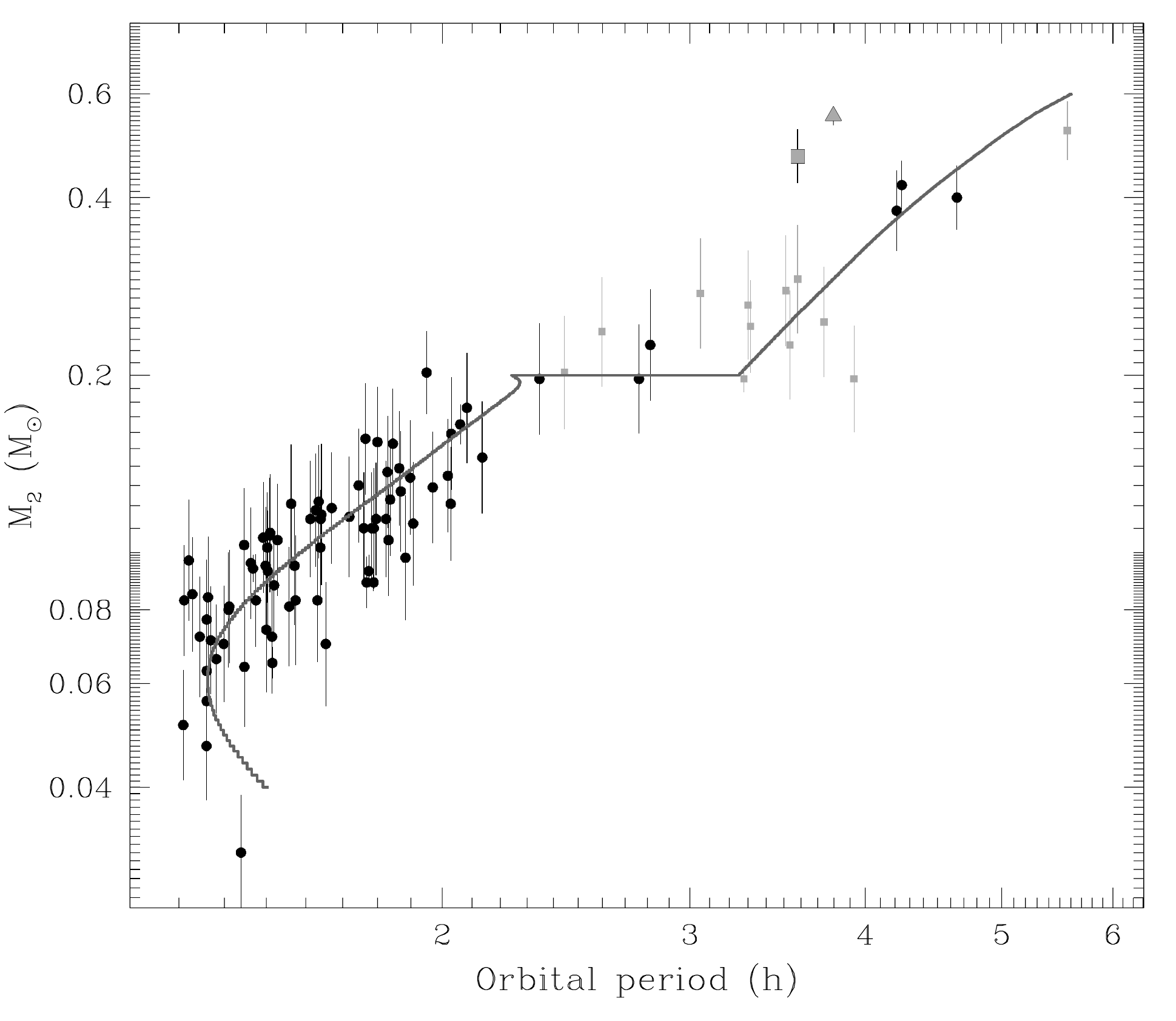}
\end{center}
\caption{Period-mass data of CV donors (data from \protect\citealt{knigge06-1}). All mass estimates come from superhump period excesses. The SW Sex systems are marked with grey squares. The large grey square corresponds to our results for \hs\, while IP Peg is plotted as a large grey triangle. The dark grey solid line shows the best-fit evolutionary model track of \protect\citet{kniggeetal11-1}.}
\label{fig_p_m2}
\end{figure}

The measured temperature and mass of the WD in \hs\ result in a time-averaged mass transfer rate of $\tmavg = (1.1 \pm 1.2) \times 10^{-9}~\Msun\ \mathrm{yr}^{-1}$. This value places \hs\ closer to the evolutionary sequences of \cite{kniggeetal11-1} than MV Lyr, TT Ari and DW UMa, but at the same time increases the scatter in $\dot{M}$ observed close to the upper edge of the period gap (fig. 5 of \citealt{townsley+gaensicke09-1}). In contrast, IP Peg (also in the 3--4 h orbital period regime) may have a time-averaged mass transfer rate as low as $\tmavg < 5 \times10^{-11}~\Msun\ \mathrm{yr}^{-1}$ as derived from its WD temperature \citep[$10\,000-15\,000$\,K;][]{copperwheatetal10-1}, almost 30 times lesser than the value inferred for \hs\ from its WD temperature. These authors also point out the possibility that the WD temperature might not be a good tracer of the medium-term mass transfer rate. The time-averaged mass transfer rate of IP Peg is also inconsistent with the results of the population synthesis study by \cite{howelletal01-1}. It is therefore clear that more direct measurements of masses and temperatures of WDs in the $3-4$ h orbital period regime are needed.       

\subsection{The M-type companion} \label{sec_discuss_md}

We have shown that \hs~contains a M-dwarf donor star with mass $M_2 = 0.47 \pm 0.05$ \Msun\ (Table~\ref{modresult}). Note that the fits presented in section~\ref{sec_tsh_model} never produced a secondary star mass smaller than 0.3~\Msun. We have approximated the radius of the donor star by its Roche-lobe volume radius \citep{eggleton83-1}. The donor radius can then be derived from its scaled value $R_2/a$ by using the binary parameters of \hs. The resulting Roche-lobe volume radius of the M-dwarf is then $R_2 = 0.43 \pm 0.03$ \Rsun. The distance to the system when this radius is taken into account in the M-dwarf spectral modelling is found to range between 740 and 940 pc for the spectral types M5 and M6, respectively. This is consistent with both the distances from the WD spectral modelling and the dynamical modelling presented in Section~\ref{sec_tsh_model}.      

\begin{figure}
\begin{center}
\includegraphics[width=\columnwidth]{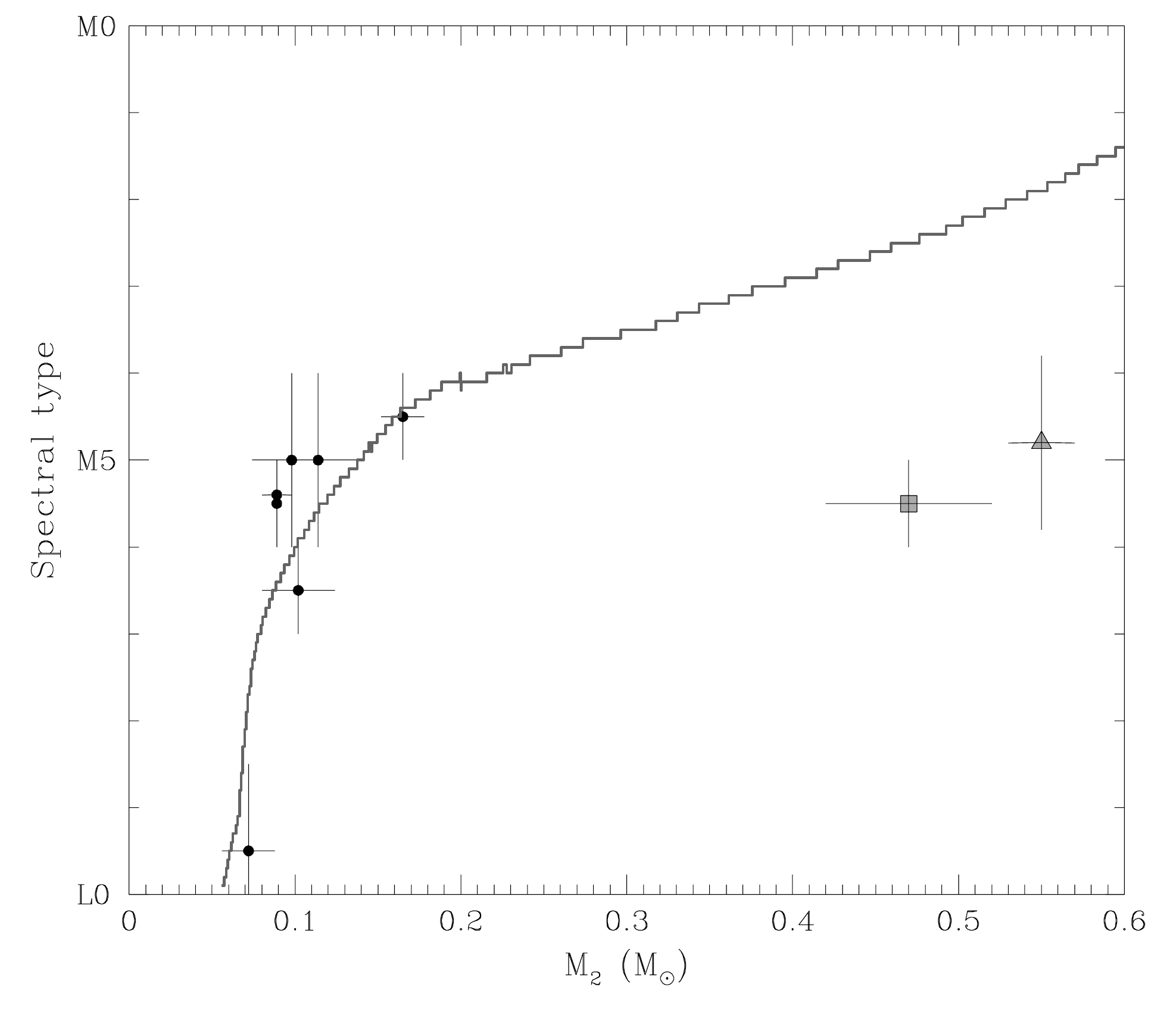}
\end{center}
\caption{Same as Fig.~\ref{fig_p_m2} but for the donor mass--spectral type data. \hs\ and IP Peg largely deviate from the evolutionary track of \protect\citet{kniggeetal11-1}.}
\label{fig_m2_sp2}
\end{figure}

However, our derived mass for the secondary star runs counter to expectations for an orbital period of 3.58 h. The semi-empirical CV evolution track of \cite{kniggeetal11-1} points to an M-dwarf mass of 0.25 \Msun. In Fig.~\ref{fig_p_m2} we show the $P$--$M_2$ diagram for CVs along with the evolutionary track of \cite{kniggeetal11-1}. Note that all mass values are not direct measurements but come from superhump period excesses, as emphasised in Section~\ref{sec_intro}. The SW Sex systems between 3 and 4 hours have been plotted as grey filled squares. We have also plotted the data for the dwarf nova IP Peg (large grey triangle), with an orbital period of 3.80 h and a secondary star mass of $0.55 \pm 0.02$\,\Msun. The secondary star in IP Peg is significantly more massive than predicted by the evolutionary track of \cite{kniggeetal11-1}.  In addition, the measured mass of the secondary star in \hs\ marginally fits in the predicted orbital period--secondary mass distribution for CVs of \cite{howelletal01-1} (cf. their fig. 5), while the secondary mass in IP Peg is clearly off.


With measured values of 0.55 \Msun\ and 0.47 \Rsun\ \cite{copperwheatetal10-1} conclude that the secondary star in IP Peg is undersized for its mass, suggesting it is in thermal equilibrium. The same might be happening in \hs\ with a secondary star mass of $0.47$ \Msun\ and a derived secondary Roche-lobe volume radius in \hs\ of 0.43 \Rsun. Taken at face value, this indicates that the donor star in \hs\ may not be bloated at all. This is not common among CVs, in which the secondary stars are believed to be significantly oversized for their mass as the result of either nuclear evolution or being out of thermal equilibrium.

In Fig.~\ref{fig_m2_sp2} we have plotted the (few) CVs with determinations of both the secondary star mass and spectral type. \hs\ is marked by a large grey square and IP Peg is plotted with a large grey triangle. Our analysis provided a secondary star spectral type of M5.5$\pm$0.5 when the donor star is observed at its inferior conjunction (i.e. phase zero). This indicates a photosphere with temperature around 2800 K \citep[see e.g.][and references therein]{bonnefoyetal14-1}. Even if \hs\ followed the semi-empirical sequence of \citet{kniggeetal11-1} and had a donor mass of 0.25 \Msun, the spectral type would still be cooler than expected. Fig.~\ref{fig_m2_sp2} also shows that the secondary star in IP Peg is underluminous for its mass. The spectral types measured for \hs\ and IP Peg are also much later than predicted by \cite{howelletal01-1} (cf. the effective temperature--orbital period distribution for CVs in their fig. 7). Again, we stress the fact that we measured the M5.5$\pm$0.5 spectral type of the secondary star in \hs\ from the average of spectra taken at orbital phases around its inferior conjunction, so any information on spectral type change along the orbit is lost. An illustrative example is the case of the polar CV ST LMi, whose secondary star shows a changing temperature with a much cooler region at orbital phase 0.8 \citep{howelletal00-1}. Therefore, more detailed studies of the secondary stars in the nova-like variables which populate the 3--4 h orbital period range during low states should be conducted in order to check how frequent these secondary stars are.

\begin{figure*}
 \psfig{angle=-90,width=17cm,file=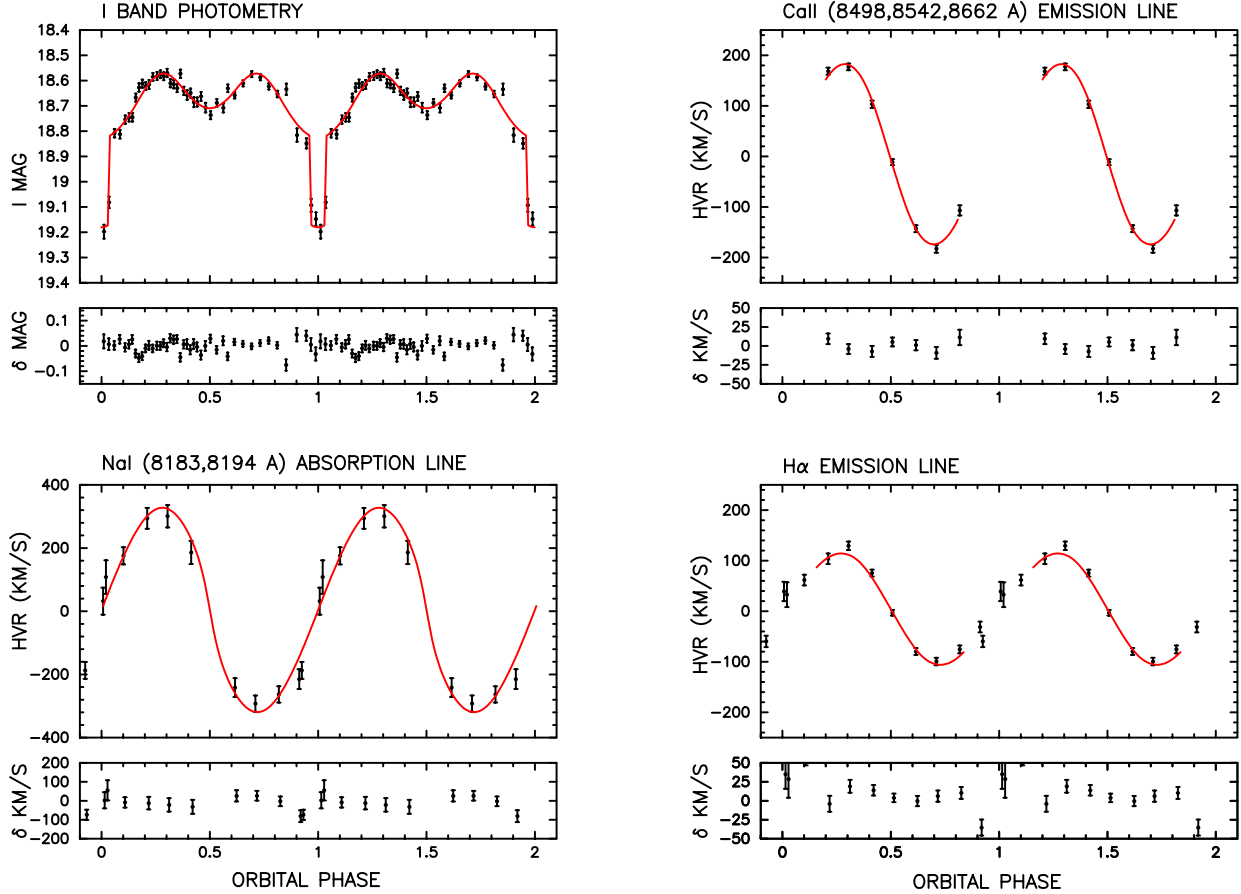}
\caption{\label{fig_fits}
From left to right and from top to bottom: NOT $I$-band ellipsoidal+WD eclipse light curve;  Ca\,{\sc ii} emission triplet radial velocity curve; Na\,{\sc i} absorption doublet radial velocity curve; H$\alpha$ emission radial velocity curve. Solid lines are the best fits with the exception of H$\alpha$ (model-predicted curve). The plots under each panel are the corresponding residuals. The orbital cycle has been repeated for clarity.}
\end{figure*}

\section{Conclusions}
We have presented the first dynamical characterisation of an eclipsing SW Sex star in the 3--4 h orbital period range during a low state (see Table~\ref{tab_sysparam}). 

We obtained time-resolved optical spectroscopy and photometry of \hs~during its 2004--2005 low state. We refined the eclipse ephemeris by including new mid-eclipse times form the photometry. Spectral modelling of the secondary star spectra taken around primary eclipse revealed a spectral type of M5.5$\pm$0.5, suggesting an effective temperature of about 2800 K. The analysis of the spectrum of \hs\ also provided the first detection of a DAB WD in a CV, with $T_\mathrm{eff} = 30\,000 \pm 5000$ K. By modelling the ULTRACAM light curve of the WD eclipse we determined solutions in the binary mass ratio--inclination angle plane. Furthermore, by simultaneously fitting the radial velocity and irradiated ellipsoidal light curves of the secondary star, we independently calculate solutions as
well as the unperturbed radial velocity amplitude of the secondary star ($K_2 = 284 \pm11$\,\kms) and the distance to the source ($d=740 \pm 26$ pc). Combining the results of the white-dwarf and secondary-star modelling we derived the stellar masses to be $M_1=0.87 \pm 0.09$ \Msun\ and $M_2=0.47 \pm 0.05$ \Msun\ for the WD and M dwarf, respectively. As in IP Peg, the M-dwarf in \hs\ seems too cool and undersized for its mass. More binary parameters from dynamical measurements in nova-like CVs with orbital periods between 3 and 4 hours during low states are needed before a critical comparison with theoretical predictions can be made. What triggers low states in CVs is still far from being fully understood.


\begin{table}
\begin{center}
\caption{\label{tab_sysparam}System parameters for \hs. $R_2$ is the volume radius of the secondary star's Roche lobe as defined by \protect\citet{eggleton83-1}.}
\def\arraystretch{1.5}
\begin{tabu} to \columnwidth {X[l]X[l]}
\hline\noalign{\smallskip}
Parameter                 & Value   \\			 
\hline\noalign{\smallskip} 
$q$  &  $0.54 \pm 0.03$\\ 
$a\ (\mathrm{R}_{\odot})$  &  $1.31 \pm 0.03$\\ 
$M_1\ (\mathrm{M}_{\odot})$  &  $0.87 \pm 0.09$\\ 
$R_1\ (\mathrm{R}_{\odot})$  &  $0.0103 \pm 0.0007$\\
$T_1\ ({\rm K})$  &  $30\,000\pm5\,000$\\ 
$M_2\ (\mathrm{M}_{\odot})$  &  $0.47 \pm 0.05$\\ 
$R_2\ (\mathrm{R}_{\odot})$  & $0.43 \pm 0.03$ \\
$K_2\ ({\rm km\,s}^{-1})$  &  $284 \pm 11$\\ 
$i\ (^{\circ})$  & $\sim 79$ \\ 
$d\ ({\rm pc})$  &   $740 \pm 26$\\
\hline\noalign{\smallskip} 
\end{tabu}
\end{center}
\end{table}

\section*{Acknowledgments}
We thank Christian Knigge and Mercedes L\'opez-Morales for enlightening discussion. We are also thankful to the anonymous referee for the helpful comments and constructive suggestions for improvement of the manuscript. The use of the {\sc molly} package developed by Tom Marsh is acknowledged. This research has been supported by the Spanish Ministry of Economy and Competitiveness (MINECO) under the grants AYA2012--38700 and AYA2010--18080. PRG acknowledges support from the MINECO under the Ram\'on y Cajal programme (RYC--2010--05762). He also acknowledges support from the XMM grant NNG05GJ22G. He is thankful to the staff of the Harvard--Smithsonian Centre for Astrophysics for a pleasant visit during which part of this work was carried out. He also wishes to thank all the M1 Group observers for their long-term dedication to the search for low states. TRM and DS acknowledge support from the UK's Science and Technology Facilities Council, grant ST/L000733/1. The research leading to these results has received funding from the European Research Council under the European Union's Seventh Framework Programme FP 2007--2013 ERC Grant Agreement n. 320964 (WDTracer). MRS acknowledges support from FONDECYT (grant 1141269) and from the Millenium Nucleus RC130007 (Chilean Ministry of Economy). VSD and ULTRACAM are supported by the STFC. Based in part on observations collected at the European Organisation for
Astronomical Research in the Southern Hemisphere, Chile, under program 074.D--0657, and on observations obtained with XMM--Newton, an ESA science
mission with instruments and contributions directly funded by
ESA Member States and NASA, and on observations made with the William Herschel Telescope
operated on the island of La Palma by the Isaac Newton Group in the
Spanish Observatorio del Roque de los Muchachos, and on observations made with the Nordic Optical Telescope operated by the Nordic Optical Telescope Scientific Association at the Observatorio del Roque de los Muchachos, La Palma, Spain, and on observations made with the IAC80 telescope operated on the island of Tenerife by the Instituto de Astrof\'\i sica de Canarias in the Spanish Observatorio del Teide. The William Herschel Telescope data were obtained as part of the 2004 International Time Programme of the night-time telescopes at the European Northern Observatory.

\footnotesize{

\bibliographystyle{mn2e}
\bibliography{/Users/prguez/Library/texmf/bib/mn-jour,/Users/prguez/Library/texmf/bib/aabib}
}

\end{document}